\documentclass{elsart}
\usepackage{epsfig}
\newcommand{\ar}{\arrowvert}

\newcommand{\da}{\dagger}

\newcommand{\cd}{\! \cdot \!}
\usepackage{amsfonts}
\usepackage{amsmath}
\usepackage{amssymb}
\usepackage{graphicx}
\newcommand{\FP}{\mathcal J}
\newcommand{\bs}{\boldsymbol}
\newcommand{\be}{\begin{equation}}
\newcommand{\ee}{\end{equation}}
\newcommand{\ba}{\begin{eqnarray}}
\newcommand{\ea}{\end{eqnarray}}

\begin{document}
\begin{frontmatter}
\hyphenation{english}
\title{Light $\bf 1^{-+}$ exotics:   molecular resonances}
\author{Ignacio J. General$^{\da 1}$,}\thanks{current address: Bayer School of Natural and Environmental  Sciences, Duquesne University, Pittsburgh, Pennsylvania 15282 }
 \author{Ping Wang$ ^\da$, Stephen R. Cotanch$
^\da$,}
\author{  Felipe J. Llanes-Estrada$ ^{\ddagger}$}
\address{$ ^\da$ Department of Physics, North Carolina State
University, Raleigh, North Carolina 27695-8202 \\
$^\ddagger$ Departamento de Fisica Teorica I, Universidad
Complutense 28040 Madrid, Spain}
\date{\today}
\maketitle
\begin{abstract}
Highlights in the search for nonconventional (non $q \bar {q}$) meson states
are
the $\pi_1(1400)$ and $\pi_1(1600)$ exotic  candidates.  Should they exist,
mounting theoretical arguments suggest that they
are tetraquark molecular resonances excitable by meson rescattering.
We report a new tetraquark calculation within a model field
theory approximation to Quantum Chromodynamics in the Coulomb gauge
supporting  this conjecture.  We also strengthen this claim
by consistently contrasting results with exotic state predictions for
hybrid ($q \bar{q} g$) mesons within the same
theoretical framework. Our findings confirm  that
molecular-like configurations involving two  color singlets (a resonance, not a bound state) are clearly favored over hybrid  or color-exotic
tetraquark meson ($q \bar{q} q \bar{q}$ atoms) formation.  Finally, to assist needed further experimental searches
we  document a  useful
off-plane correlator for establishing the
structure of these exotic systems along with similar, but anticipated much narrower, states that should exist
in the charmonium and bottomonium spectra.

\end{abstract}

\scriptsize{
PACS: 12.39.Mk; 12.39.Pn; 12.39.Ki;  12.40.Yx

{\it{Keywords}}:  Exotic mesons, tetraquarks, QCD Coulomb gauge, effective Hamiltonian}
\end{frontmatter}





\hspace*{\parindent}

The existence and understanding of exotic (non $q \bar q$ and $qqq$) hadrons is
one of the few remaining closures to the standard model.  Such states are expected
according to Quantum Chromodynamics (QCD) and are of intense experimental interest.
In the light quark sector, there are two solid isovector
candidates at 1.3-1.4 and  1.6 GeV, $\pi_1(1400)$ and $\pi_1(1600)$, each having $J^{PC} = 1^{-+}$
 \cite{Adams:1998ff,Chung:1999we}.
Although  doubts exist about the  1.4 GeV
candidate \cite{Dzierba:2005jg}, a new analysis   \cite{Adams:2006sa} supports it.
The signature is observed as a p-wave resonance in the $\eta \pi_0$ system and it therefore has odd parity $P$ but
even  charge conjugation $C$ yielding the quantum numbers
$J^{PC}=1^{-+}$.
Since a $q\bar{q}$ state with orbital $L$ and spin $S$ coupled to $J = 1$
must have $P=(-1)^{L+1}$ and $C=(-1)^{L+S}$, this state is clearly
exotic.

\begin{figure}[b]
    \includegraphics[width=1.0\textwidth]{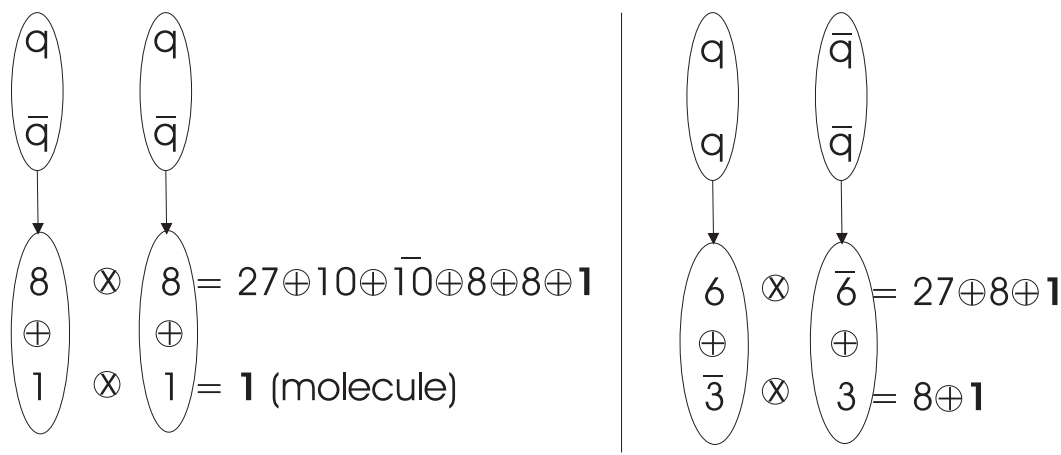}
\caption{Four independent tetraquark color  schemes. One is a  singlet-singlet molecule while the other three are more exotic atoms (octet and two diquark schemes).}
\label{fig:colorwf}
\end{figure}

Assuming these states exist,  the theoretical situation is even more controversial.  The debate
concerning their structure is among four possible scenarios:  1) a hybrid ($q \bar {q} g$) meson; 2) a tetraquark atom
($q \bar {q} q \bar {q}$ involving intermediate color states that are not singlets); 3) a tetraquark molecular bound state
of two conventional mesons; 4) a tetraquark molecular resonance
($q \bar {q} q \bar {q}$ involving two intermediate color states that are  singlets but not observed mesons).  All four scenarios can produce  $J^{PC}=1^{-+}$ states but  the first two are more exotic since the tetraquark molecule
is color equivalent to a conventional meson-meson two-body state (see Fig. \ref{fig:colorwf}).

Lattice results, now  performed with more realistic  lower quark masses
\cite{Hedditch:2005zf}, have focused upon the hybrid scenario but
find that hybrid correlators can only  produce a
state as low as 1.9 to 2.1 GeV.
A more recent lattice calculation
\cite{Cook:2006hh}  claims to find two hybrid meson masses below 2 GeV,
however they use an ad-hoc extrapolation that is quadratic in the pion mass which
increases uncertainties.  Similarly, the lightest Flux Tube model $1^{-+}$ predictions~\cite{Barnes,Close,Katja} are
also near 2 GeV spanning the region of 1.8 to 2.1 GeV.
This is consistent with agreement among other model approaches, based on either the
concept of constituent gluons \cite{Iddir:2005xm} or  field-theory
calculations (see below) generating mass-gaps
\cite{Llanes-Estrada:2000hj}, that the lightest hybrid mesons with just one
constituent gluon should be somewhat heavier than the 1.4, 1.6 GeV
experimental candidates. Finally, both well-established spin \cite{Iddir:1988jd}  and flavor
 \cite{Chung:2002fz} selection
rules indicate that the above mentioned $\eta\pi$ signature cannot
be due to a hybrid meson decay. Therefore it would appear that the two $\pi_1$ states can not be
theoretically explained as hybrid mesons.


In search of other
explanations,
a potential model lifetime calculation
\cite{Zhang:2001sb} has  ruled out a molecular   $\eta(1295)\pi$ or
$\eta(1440)\pi$ bound state.  However,
it has been shown that meson-rescattering, specifically
in the $\eta  \pi$ channel, could produce a resonance with
this signature  \cite{Bass:2001zs,Szczepaniak:2003vg}. Also,
 it has been suggested \cite{Donnachie:1998ty}  that the 1.6 GeV resonance could
be interfering with a background to produce the 1.4 GeV structure.

Summarizing the status of this situation,  while the $\pi_1(1600)$ can not be firmly
precluded as a hybrid meson or exotic
tetraquark, the $\pi_1(1400)$  seems explainable only as a molecular resonance excited by
meson rescattering.  The purpose of this work is to confirm the latter  by a new theoretical
analysis which also predicts that the  $\pi_1(1600)$ is not a color exotic system.
Our formalism, referred to as the Coulomb Gauge Model (CGM), has  been successfully
established in both the quark~\cite{Llanes-Estrada:1999uh,Llanes-Estrada:2001kr,Llanes-Estrada:2003wr} and gluon~\cite{LBC} sectors   and is based upon the
exact QCD Hamiltonian in the Coulomb gauge \cite{T-D-Lee} given by
\begin{eqnarray}
  H_{\rm QCD} &=& H_q + H_g +H_{qg} + H_{C}    \\
H_q &=& \int d{\bs x} \Psi^\dagger ({\bs x}) [ -i
{\mbox{\boldmath$\alpha$\unboldmath}} \cdot
{\mbox{\boldmath$\nabla$\unboldmath}}
+  \beta m] \Psi ({\bs x})   \\
 H_g &=& \frac{1}{2} \int \!\! d {\bs x} \!\! \left[
\FP^{-1}{\bf \Pi}^a({\bs x})\cdot \!\!  \FP {\bf
\Pi}^a({\bs x}) +{\bf B}^a({\bs x})\cdot{\bf B}^a({\bs x}) \right] \\
H_{qg} &=&  g \int d {\bs x} \; {\bf J}^a ({\bs x})
\cdot {\bf A}^a({\bs x}) \\
H_C &=& -\frac{g^2}{2} \int d{\bs x} d{\bs y} \rho^a ({\bs x}) \FP^{-1}
K^{ab}( {\bs x},{\bs y}  ) \FP
 \rho^b ({\bs y})   \ .
 \label{model}
\end{eqnarray}
Here $g$ is the QCD coupling,
$\Psi$  the quark field   with current quark
mass $m$,
${\bf A}^a$  the gluon fields
satisfying the transverse gauge condition,
$\mbox{\boldmath$\nabla$\unboldmath}$ $\cdot$ ${\bf A}^a = 0$, $a
= 1, 2, ... 8$, ${\bf \Pi}^a $ the conjugate fields and ${\bf
B}^a$  the non-abelian magnetic fields,
${\bf B}^a = \nabla \times {\bf A}^a + \frac{1}{2} g f^{abc} {\bf A}^b
\times {\bf A}^c  $.
The color densities, $\rho^a({\bs x})= \Psi^\dagger({\bs x}) T^a\Psi({\bs x}) +f^{abc}{\bf
A}^b({\bs x})\cdot{\bf \Pi}^c({\bs x})$, and quark color currents, ${\bf
J}^a= \Psi^\dagger ({\bs x})
\mbox{\boldmath$\alpha$\unboldmath}T^a \Psi ({\bs x})$, entail the $SU_c(3)$ color matrices,  $T^a = \frac{\lambda^a}{2}$, and structure constants, $f^{abc}$.
The
Faddeev-Popov determinant, $\FP = {\rm det}(\mathcal M)$, of the
matrix ${\mathcal M} = {\mbox{\boldmath$\nabla$\unboldmath}} \cdot
{\bf D}$ with covariant derivative ${\bf D}^{ab} =
\delta^{ab}{\mbox{\boldmath$\nabla$\unboldmath}}  - g f^{abc} {\bf
A}^c$, is a measure of the gauge manifold curvature and the kernel
in Eq. (5) is given by $K^{ab}({\bs x}, {\bs y}) = \langle{\bs x},
a|{\mathcal M}^{-1} \nabla^2 {\mathcal M}^{-1}  |{\bs y}, b\rangle$. The
Coulomb gauge Hamiltonian is renormalizable, permits resolution of
the Gribov problem, preserves rotational invariance, avoids
spurious retardation corrections, aids identification of dominant,
low energy  potentials and does not introduce unphysical degrees
of freedom (ghosts) \cite{dz}.
To make the problem tractable,  the
Coulomb instantaneous kernel is approximated by its vacuum expectation value,
yielding an effective potential field theory, $H_{\rm QCD} \rightarrow H^{\rm eff}_{QCD}$
\be
H_C \rightarrow H^{\rm eff}_C =  -\frac{1}{2} \int d{\bs x} d{\bs y} \rho^a ({\bs x})
\hat{V}(\ar {\bs x}-{\bs y}
\ar ) \rho^a ({\bs y}) \ ,
\ee
with confinement  described by a Cornell potential,
$\hat{V}(r)=-\frac{\alpha_s}{r}+\sigma r$, where  the string tension,
$\sigma=0.135$ GeV$^{2}$, and $\alpha_s=0.4$  have been independently
determined from conventional meson studies
\cite{Llanes-Estrada:1999uh,Llanes-Estrada:2001kr,Llanes-Estrada:2003wr}
within the same field theory  approach.
We also use the lowest order, unit value, for the Faddeev-Popov determinant in the $H_g$ term and treat the $H_{qg}$ interaction
using perturbation theory.
Lattice data confirms the Cornell potential form  between
static sources \cite{Bali:1997am} and further provides  the
scale of the gluon mass gap \cite{Morningstar:1999rf}, that is needed
to fit a counterterm in the gluon gap equation. The
remaining parameters are the reasonably well known current quark masses at some high energy scale
where the mass function runs perturbatively.
The quark sector
Hamiltonian then takes the form of the Cornell coupled-channel model
\cite{Eichten:1978tg}.
We note that  three-body forces
\cite{Szczepaniak:2006nx,Dmitrasinovic:2004cu} are omitted, however
based upon successful three-body applications \cite{LBC} we submit
the CGM should capture the dominant features of  a multi-parton
spectrum.

Before presenting our tetraquark results we highlight our recent hybrid meson calculation
\cite{General:2006ed}. Since the gluon carries a color octet charge, the
quark and antiquark are also in a color octet wavefunction with elements
$T^a_{ij}$
(they repel each other at short distance).
In the hybrid rest frame
there are two independent
three-momenta ${{\bs q}_+ =  \frac{{\bs q}+\overline{\bs q}}{2}}$,
${\bs q_- =  \bs q- \overline{\bs q} }$ and one dependent ${\boldsymbol g =
-\bs q -\overline{\bs q} }=-2{\bs q}_+ $.
The leading hybrid Fock space wavefunction can  therefore be constructed from the respective quark, anti-quark and gluon quasiparticle operators
$B^ \dag_{\lambda_1{\mathcal
C}_1}(\boldsymbol{q}),
 D^ \dag_{\lambda_2{\mathcal C}_2}(\overline{\boldsymbol{q}})$ and
$\alpha^{a\dag}_{\mu}({ \boldsymbol g})$
  \be
|\Psi^{JPC}\rangle = \int \!\! \int \!\!
\frac{d\boldsymbol{q}_+}{(2\pi)^3}
\frac{d\boldsymbol{q}_-}{(2\pi)^3} \Phi^{JPC}_{\lambda_1 \lambda_2
\mu}(\boldsymbol{q}_+,\boldsymbol{q}_-)
T_{{\mathcal C}_1{\mathcal C}_2}^a B^ \dag_{\lambda_1{\mathcal
C}_1}(\boldsymbol{q})
 D^ \dag_{\lambda_2{\mathcal C}_2}(\overline{\boldsymbol{q}})
\alpha^{a\dag}_{\mu}({ \boldsymbol g}) |\Omega \rangle \ .
\ee
An angular momentum expansion for the lightest $1^{-+}$ state reveals a
required   p-wave excitation in one of the two orbital wave functions.  Consult Ref.
\cite{General:2006ed} for  complete details. Significantly, in agreement
with earlier findings \cite{Llanes-Estrada:2000hj}, the predicted hybrid masses are
about 2 GeV for the ground state (parity +) quadruplet and
2.2 and 2.4 GeV, respectively, for the first $1^{-+}$ exotics.
The repulsive nature of the short range $q \bar{q}$ interaction
potential kernel and the large
gluon mass gap are responsible for these large masses.
We also performed a parameter sensitivity and error analysis study and concluded there was no model possibility
to lower one of these states  near the 1.6 GeV
candidate and therefore rule out this, and even more clearly the 1.4 GeV, state as a hybrid.

Returning to our thrust, we report  results for $qq\bar{q}\bar{q}$
spectroscopy (see also a preliminary study  \cite{Cotanch:2006ed}). This system was first investigated  in the bag model
\cite{Jaffe:1976ig} and then more extensively by Ref. \cite{Aerts:1979hn} with subsequent
potential model applications  reported by Refs.
\cite{Badalian:1987jg,Maiani}.  While these studies have some similarity to our model,
we submit our results are more robust since our approach is much more comprehensive,
has many QCD elements with
no new parameters to be determined and employs a realistic potential kernel  extracted
from lattice gauge theory.
The leading tetraquark Fock space wavefunction  is \cite{Llanes-Estrada:2003th,Santopinto:2006my}
 \ba
    |\Psi^{JPC}\rangle = \int \!\!\!\! \int \!\!\!\! \int \!\!
    \frac{d\boldsymbol{q}_A}{(2\pi)^3}
\frac{d\boldsymbol{q}_B}{(2\pi)^3}
    \frac{d\boldsymbol{q}_I}{(2\pi)^3} \Phi^{JPC}_{\lambda_1 \lambda_2
    \lambda_3
\lambda_4}(\boldsymbol{q}_A,\boldsymbol{q}_B,\boldsymbol{q}_I) \times   \nonumber \\
    R^{{\mathcal C}_1{\mathcal C}_2}_{{\mathcal C}_3{\mathcal C}_4}
    B^{\dag}_{\lambda_1{\mathcal C}_1}(\boldsymbol{q}_1)
    D^{\dag}_{\lambda_2{\mathcal C}_2}(\boldsymbol{q}_2)
    B^{\dag}_{\lambda_3{\mathcal C}_3}(\boldsymbol{q}_3)
    D^{\dag}_{\lambda_4{\mathcal C}_4}(\boldsymbol{q}_4)|\Omega \rangle .
  \ea
In the $cm$ there are three independent
 momenta which we take to be ${{\bs q}_A =  \frac{{\bs q}_1 - {\bs q}_2}{2}}$,
 ${{\bs q}_B =  \frac{{\bs q}_3 - {\bs q}_4}{2}}$  and
 ${{\bs q}_I = \frac{{\bs q}_3 + {\bs q}_4}{2} -  \frac{{\bs q}_1+{\bs q}_2}{2}}$   (${\bs q_1}$, ${\bs q_3}$ for the quarks and
${\bs q_2}$, ${\bs q_4}$ for the anti-quarks).  The color
 matrices $R^{{\mathcal C}_1{\mathcal C}_2}_{{\mathcal C}_3{\mathcal C}_4}$
yielding color-singlet wavefunctions follow from the $SU_c(3)$ algebra
depicted  in Fig. \ref{fig:colorwf}.
The radial part of $\Phi^{JPC}_{\lambda_1 \lambda_2 \lambda_3 \lambda_4}
(\boldsymbol{q}_A,\boldsymbol{q}_B,\boldsymbol{q}_I)$
is chosen to be a gaussian, $exp(-\frac{q^2_A}{\alpha^2_A} - \frac{q^2_B}{\alpha^2_B} -
\frac{q^2_I}{\alpha^2_I})$, with variational parameters $\alpha_A,
\alpha_B$ and $ \alpha_I$
for s-wave states and a gaussian multiplied by
$q^2_i/\alpha^2_i\;\;(i=A,B,I)$ corresponding to orbital $L_i = 1$, when
treating p-wave states.
Note, as for hybrid mesons, the
ground state tetraquark multiplet has positive parity  and
that constructing $1^{-+}$ exotics requires one of the three orbitals to be a
p-wave.
Using the variational principle, the tetraquark mass is then given by
\begin{eqnarray}
\!\!\!\!\!\!M_{J^{PC}} \leq \frac{\langle\Psi^{JPC}|H^{\rm
eff}_{QCD}|\Psi^{JPC}\rangle} {\langle\Psi^{JPC}|\Psi^{JPC}\rangle}
= M_{self}+M_{qq}+M_{\bar{q} \bar{q}}+M_{q \bar{q}}+M_{annih} \ .
\end{eqnarray}
\begin{figure} [b]
    \hspace{1.5cm}
    \includegraphics[width=.8\textwidth]{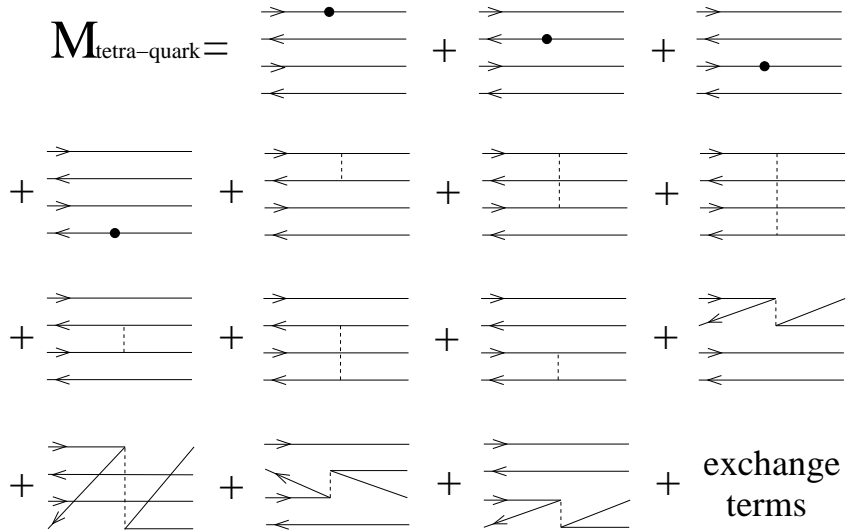}
\caption{Equal-time diagrams for the expectation value of the
model Hamiltonian.}
\label{fig:Fdiagrams}
\end{figure}
Contributions to the Hamiltonian expectation value are summarized in Fig. \ref{fig:Fdiagrams} and correspond to
4 self-energy, 6 scattering, 4
annihilation and 70 exchange terms, each of which can be reduced to 12 dimensional integrals that are evaluated in momentum space. Because of the computationally intensive nature of this analysis, the
hyperfine interaction was not included.
Complete expressions will be given in another publication, but as an
example note the annihilation contribution, not possible in standard quark
models, is
\begin{eqnarray}
&& M_{annih}=\int\!\!\!\int\!\!\!\int\!\!\!\int\!
\frac{d\boldsymbol{q}_1 d\boldsymbol{q}_2 d\boldsymbol{q}_3 d\boldsymbol{k}}{(2\pi)^{12}}
V(\boldsymbol{q}_1+\boldsymbol{q}_2)
u_{\lambda_1^{'}}^\dag({\boldsymbol q}_1+\boldsymbol{k})v_{\lambda_2^{'}}({\boldsymbol q}_2-\boldsymbol{k}) \times \nonumber \\ &&
\; \; \; \; \; \; \; v_{\lambda_2}^\dag({\boldsymbol q}_2)
u_{\lambda_1}({\boldsymbol q}_1)
 \Phi_{\lambda_1\lambda_2
\lambda_3\lambda_4}^{JPC\dag}({\boldsymbol q}_1,{\boldsymbol q}_2,{\boldsymbol q}_3)\Phi_{\lambda^{'}_1
\lambda^{'}_2 \lambda_3\lambda_4}^{JPC}({\boldsymbol q}_1+\boldsymbol{k},{\boldsymbol q}_2-\boldsymbol{k},{\boldsymbol q}_3) ,
\end{eqnarray}
involving Dirac spinors $u_{\lambda_1}$ and $v_{\lambda_2}$.
This raises the mass of the  isoscalar states
relative to the states with higher isospin, which  would otherwise
typically be heavier due to the exclusion principle applied to
equal-flavor quarks.

\begin{figure} [b]
   \includegraphics[width=1.0\textwidth]{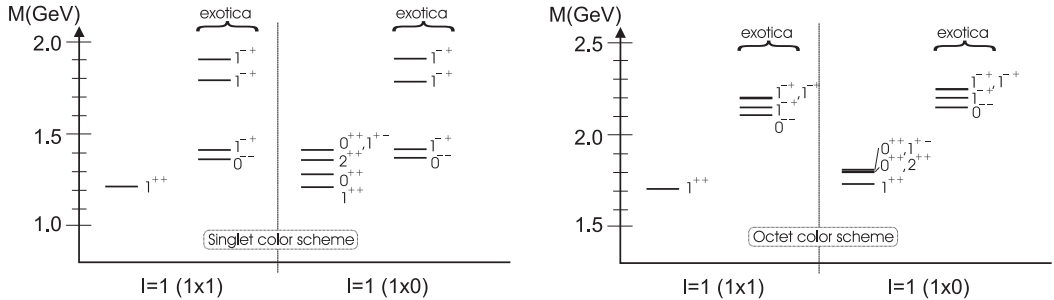}
   \caption{Tetraquark   $I = $ 1 singlet (molecule) and octet (atom) schemes  spectra.}
   \label{4qmolecule}
\end{figure}

Performing large-scale Monte Carlo calculations (typically 50 million samples), has conclusively determined that the molecular representation (i.e. singlet-singlet) produces the lightest mass  for a given $J^{PC}$. This is due to suppression of certain interactions in this color scheme from vanishing color factors for every parton pair  which does not occur in the other representations. Also, there are additional, repulsive forces in the
more exotic  color schemes.
Using $m_u = m_d = 5$ MeV,  the predicted tetraquark ground state is the
non-exotic vector $1^{++}$ state in the molecular representation
with  mass around 1.2 GeV.   Figure \ref{4qmolecule}
depicts the predicted tetraquark spectra for states having conventional
and exotic quantum numbers in both singlet and octet color representations.
The quark annihilation interactions ($q \bar{q} \rightarrow
g \rightarrow  q \bar{q}$) in the $I_{q \bar{q}} = 0$ channel
generate isospin splitting contributions, up to several hundred MeV, in the octet
but not singlet scheme as slightly  illustrated in the figure. Isospin splitting is a consequence
of a more proper field theory treatment, not present in conventional quark models.  The annihilation interaction terms are repulsive, yielding octet states with $I =2$  lower than the $I = 1$ which are lower than the $I = 0$. This is intuitively contrary to expectations that I= 2 states are higher based upon the Pauli principle that identical quarks repel. The molecular states are all  isospin degenerate and
the lightest exotic molecule  is an intriguing  $0^{--}$ with
mass 1.35 GeV that could be detected in
a sophisticated p-wave analysis of an $\omega\pi$ spectrum.  Because of the isospin degeneracy, there will be several molecular
tetraquark states with the same $J^{PC}$ in the 1 to 2 GeV region.  Further, these states can be observed in different electric charge channels (different $I_z$) at about the same energy, which is
a useful experimental signature.
The
lightest $1^{-+}$ is predicted near 1.4 GeV which is close to the observed $\pi(1400)$, suggesting this state has a molecular resonance structure.
The computed  mass for $1^{-+}$ states with  more exotic octet color configurations are all above 2 GeV.  This is consistent with  model predictions
\cite{General:2006ed} for exotic hybrid meson ($q \bar{q} g$) $1^{-+}$ states also lying above 2 GeV due to repulsive color octet quark
interactions.  Finally,  for any $J^{PC}$ state, including
the $1^{-+}$, the computed masses (not shown) in both the triplet  and the sextet diquark color representations are all heavier than in the singlet representation and  comparable to the octet
scheme results.
Our predictions for the lightest
exotic molecules are given in Table \ref{table:spectrum}.


\begin{table}[t]
\caption{Selected tetraquark molecular exotic states. For  $q\bar{q}$  pairs, $S_i$
and $L_i$, $i = A,B$, are the total spin and orbital angular
momentum  and $L_I$  is the
orbital angular momentum between the pairs. The variables not shown are assumed to be 0.
Units are GeV.
} \label{table:spectrum}
\begin{tabular}{|c|c|c|c|c|} \hline
$(u\bar{u})_1(u\bar{u})_1$  &I=2& I=1(1x1)&I=1(1x0)&I=0(1x1$\simeq$ 0x0)
\\
\hline
$1^{-+}$ ($L_I=1$)                   & - & 1.42 & 1.42 & - \\
$1^{-+}$ ($L_A=S_A=1$)& 1.80 & 1.80 & 1.80&1.80
\\
$1^{-+}$ ($L_A=S_B=1$)& 1.91 & 1.91 & 1.91&1.91
\\
$0^{--}$ ($L_I=S_A=1$)& 1.35 & 1.36 & 1.36&1.36
\\
\hline
\end{tabular}
\end{table}

\begin{figure} [b]
    \hspace{2cm}
    \includegraphics[width=.55\textwidth,angle=-90]{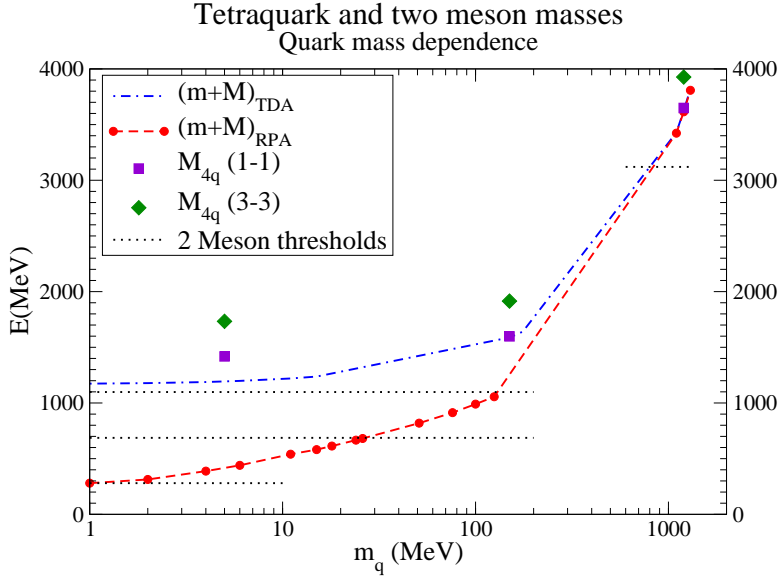}
    \caption{Dotted lines, from bottom to top, are the
$\pi\pi$, $\pi\eta$, $\pi\eta'$,$\pi \eta_c$ observed thresholds, respectively. The two other
plots are the pseudoscalar $m_\pi + M_{q\bar{q}}$ model predictions in the (chiral respecting)
RPA and (chiral violating) TDA. The squares and diamonds correspond respectively  to predicted scalar
isoscalar tetraquark masses in
the color singlet-singlet   and  triplet-triplet
schemes.
Note the two-meson
calculation \cite{Llanes-Estrada:2001kr} has $\sigma=$ 0.18
Gev$^2$, $\alpha_s=0$, whereas our results use $\sigma=$ 0.135
Gev$^2$, $\alpha_s=0.4$, but the difference among both sets is known to
be small for the ground state.}
 \label{fig:massdep}
\end{figure}

It is interesting to document the current quark mass dependence of our results.
This is illustrated in Fig. \ref{fig:massdep}, where the calculated rest mass corresponding to
different four quark systems (tetraquarks and two mesons) is displayed
with one $q\bar{q}$ pair being held fixed at a low 1-5
MeV while the other pair varies from 1 to 1300 MeV. Note the predicted  $c\bar{c}u\bar{u}$  tetraquark mass
is relatively closer to the two meson
decay threshold  indicating a narrow  decay width to the  $\pi \eta_c$ and  $\eta \eta_c $ channels
(more so for $b\bar{b}$, not displayed). Also note a
quark-meson exchange model \cite{Vijande:2003ki} claims some $J=1$
partners will be absolutely stable.
The lowest-lying hidden-charm exotics are predicted
in Table \ref{table:cc} and can be searched for in hidden
charm decays (p-wave $\eta_c \pi_0$, $\eta_c \eta $ for example).

\begin{table} [t]
\caption{
Predicted masses of hidden-charm exotic $1^{-+}$ mesons (molecular configuration).}
\label{table:cc}
\begin{tabular}{|c|c|c|}
\hline
     $m_c = 1.2$ GeV          & $m_q$     & $M_{c\bar{c}q\bar{q}}$ \\
\hline
$c\bar{c}q\bar{q}$ &  0        & 4.04 GeV \\
$c\bar{c}q\bar{q}$ &  110 MeV  & 4.10 GeV \\
$c\bar{c}q\bar{q}$ &  150 MeV  & 4.15 GeV \\
\hline
\end{tabular}
\end{table}


\begin{figure} [b]
    \hspace{1cm}
    \includegraphics[width=.75\textwidth]{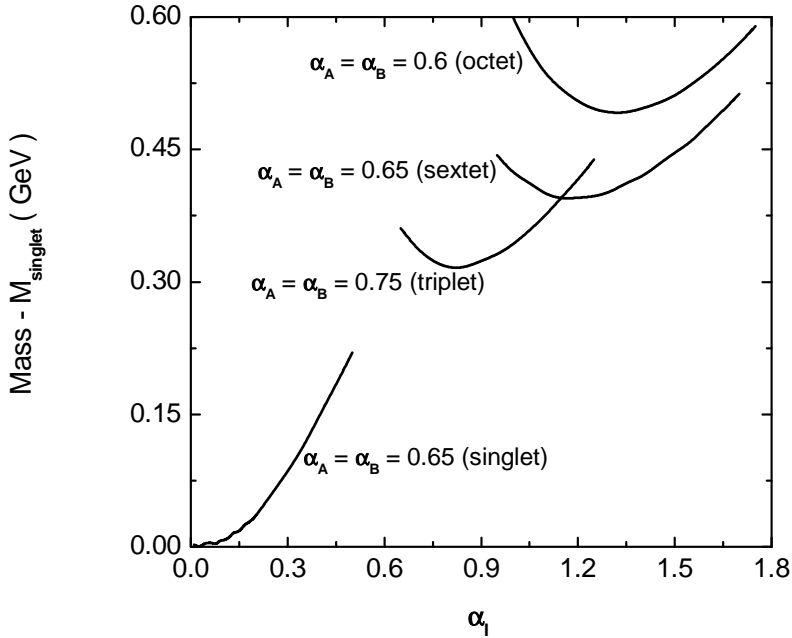}
    \caption{
Tetraquark mass of the ground state isoscalar $0^{++}$ as a
function of the intercluster variational parameter, for
various color configurations. Note  the more exotic color configurations are above
1.7 GeV.}
    \label{fig:schemes}
\end{figure}

Figure \ref{fig:schemes} displays the sensitivity of the scalar tetraquark mass to the variational parameters for different
color schemes. The color singlet-singlet configuration, at 1.28 GeV, is clearly  the lightest,  in agreement with previous predictions
\cite{Weinstein:1990gu}.
Similar results are
obtained for other $J^{PC}$ configurations.
This suggests that the lightest
 tetraquark states, including the $\pi_1(1400)$ should it
exist, are best interpreted as  molecules of color singlets (the ``extraordinary hadrons'' \cite{Jaffe:2007id}) and directly excitable via meson rescattering.

\begin{figure}
    \hspace{-0.5cm}
    \includegraphics[width=.5\textwidth]{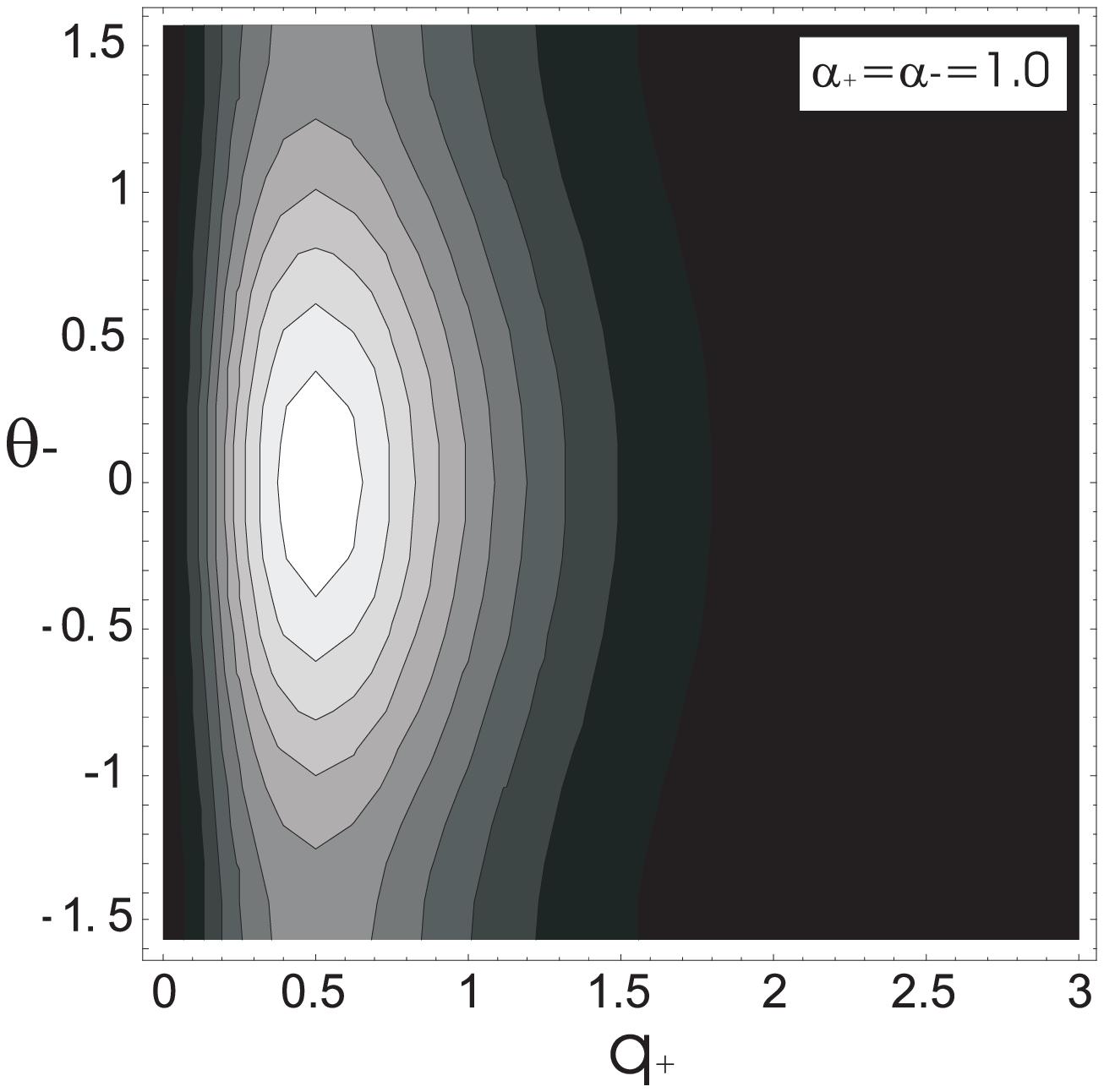}
    \includegraphics[width=.5\textwidth]{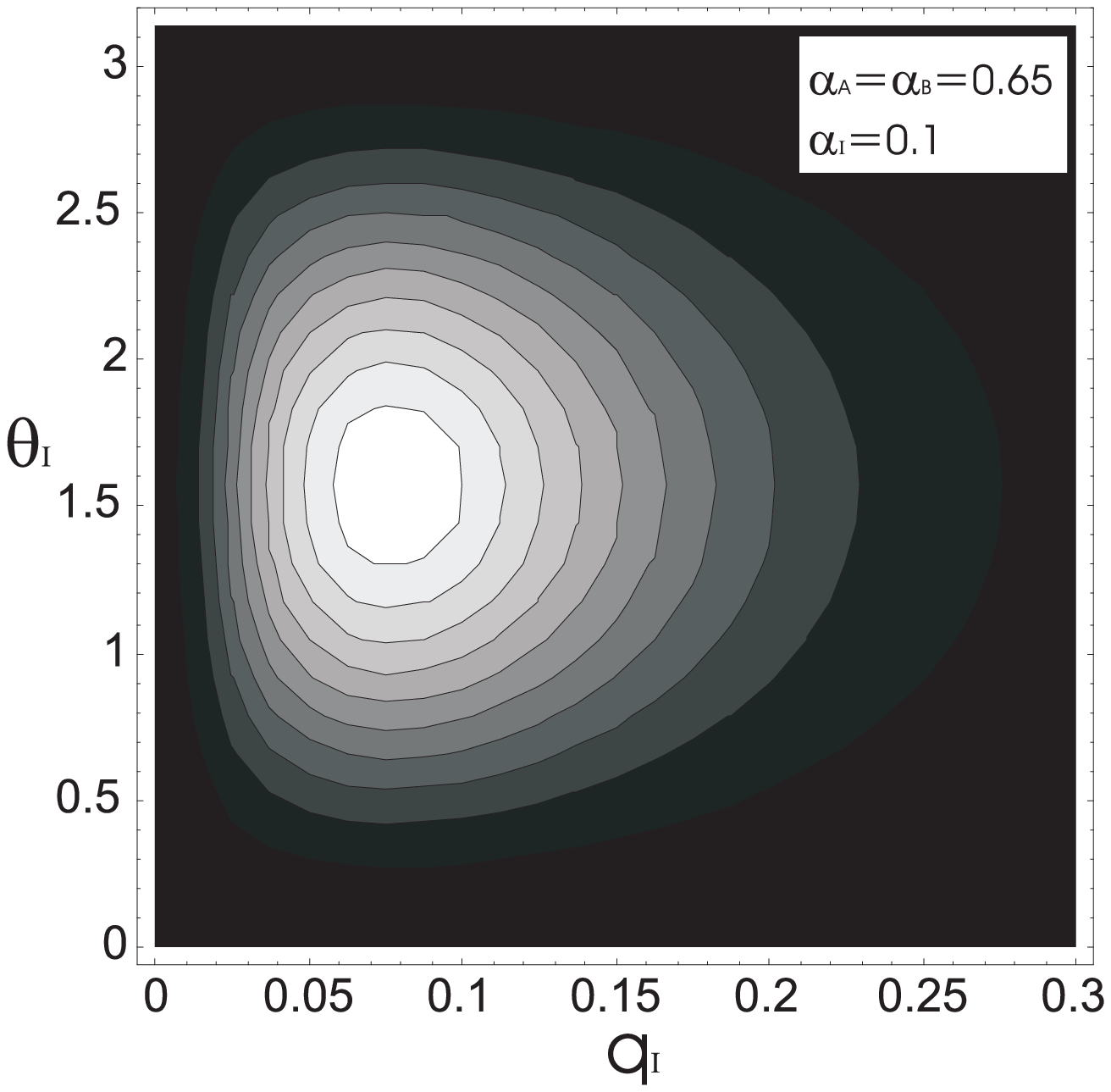}
\caption{Hybrid  (left) and tetraquark (right) meson probability density contour plots. The  left horizontal axis is half the momentum of
the constituent gluon, $q_+ = g/2$. The left vertical axis is
the angle between the $q \bar{q}$ relative momentum ${\bs q}_-$ and the hybrid total angular momenta $\bs J$.  The
right horizontal axis corresponds to the intracluster momentum $q_I$ and the  angle between
this momentum and $\bs J$ for the tetraquark. Angles in radians, momenta in GeV.}
\label{fig:contourplots}
\end{figure}


For further dynamical insight,  Fig.  \ref{fig:contourplots} details
contour plots of the probability densities using the hybrid (left) and tetraquark (right) variational
wavefunctions.
Note the depletion of the wavefunction at
low-momentum,  reflecting the rising confining potential at large
distance, and the significantly different parton momentum distributions  between hybrids
and tetraquark systems.  This difference in momentum distributions will produce distinct decay
signatures which can be utilized to identify hybrids and tetraquarks as we now
discuss.

\begin{figure}
    \hspace{1.75cm}
    \includegraphics[width=.7\textwidth]{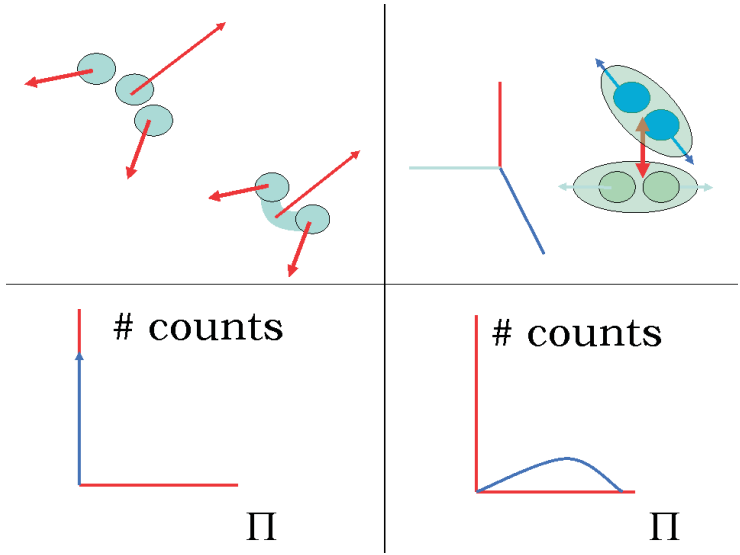}
       \caption{Differences between internal in-plane three-body hybrid (top left) and off-plane tetraquark (top right) meson decays. From the Frank-Condon principle  the internal momentum
distribution is reflected in the momentum distribution of the final
products yielding an off-plane
final state momentum observable, $\Pi$, that is very different for
$q\bar{q}g$ (bottom left) and $qq\bar{q}\bar{q}$ (bottom right) mesons.
 } \label{fig:offplanes} \end{figure}

It has been proposed \cite{Close:1994hc} that $1^{-+}$
exotic hybrid mesons decay preferentially to a meson pair in a relative
s-wave,
where one of the mesons is a p-wave (axial) meson. However that
prediction was based on the Flux Tube model and not known to hold exactly
in any limit.
We have recently \cite{General:2006ed} detailed a less model dependent
decay signature based upon
the Franck-Condon principle of molecular physics, which predicts
that the
momentum distribution of decay products parallels the internal momentum
distribution of the parent meson. This is an exact statement in the
infinitely heavy quark limit, however it is also
 useful for physical quarks. Our previous application
\cite{General:2006ed} was designed to distinguish $q\bar{q}$ from
$q\bar{q}g$ mesons and is now generalized to the four-body
problem. For two and three-body systems, in the $cm$ partons'
momenta are restricted to a plane, however for tetraquarks there
are off-plane degrees of freedom which can be exploited as an
identification signature (see the cartoon in Fig.
\ref{fig:offplanes}).
 One observable sensitive to this is an exclusive measurement of decay to four mesons.
By then determining their momenta ${\bs p}_{i = 1,2,3,4}$ in the  $cm$ of the exotic candidate,  kinematic cuts can be applied for every
two and  three meson groups to eliminate intermediate two and three-body
resonances that could confuse the analysis. From the pure (and
much reduced) four-body decay sample count,
 one can construct  the off-plane correlator, $\Pi$, for any three
momenta  which is the volume of the parallelepiped they form in
momentum space, \be \Pi({\bs p}_1,{\bs p}_2,{\bs p}_3,{\bs p}_4) =
\frac{( ({\bs p}_1\times {\bs p}_2) \cd {\bs p}_3)^2}{\sqrt{ \ar
{\bs p}_1 \times {\bs p}_2 \ar \ar {\bs p}_2 \times {\bs p}_3 \ar
\ar {\bs p}_1 \times {\bs p}_3 \ar \ar {\bs p}_1 \times {\bs p}_4
\ar \ar {\bs p}_2 \times {\bs p}_4 \ar \ar {\bs p}_3 \times {\bs
p}_4 \ar
}} \ , \nonumber \\
\ee
which has been  normalized  by the area of the six faces.
This dimensionless off-plane correlator is
useful for the very different scales involved in light and
heavy quark physics, positive definite and  invariant under
permutation of the four momenta. In the $cm$ system only
three momenta are linearly independent. Taking them equal and along the
edges of a cube yields $\Pi=8^{-1/4}=0.59$, with a more general
maximum value  close to 0.707. Therefore
the value of the correlator event by event will be a random variable
distributed between 0 and 0.707 and, with sufficient statistics, one
could deduce the
internal structure of the decaying resonance (bottom right plot of Fig. \ref{fig:offplanes}).
This could then be compared to four-meson decays of  established conventional and hybrid meson benchmarks, as distinguished
using the procedure outlined in Ref. \cite{General:2006ed},
 which would be radically different (their $\Pi$ volumes collapse to 0 in the heavy quark limit).
Finally, we are currently calculating tetraquark decay widths to compare with both conventional and
hybrid meson decays to two and multi-meson final states.  These  will also reflect hadronic structure
differences and therefore aid exotic identification.
Results will be reported in a future communication.


In conclusion, we submit that the observed $1^{-+}$ exotics
below 2 GeV are not color exotic hadrons but rather somewhat more conventional tetraquark molecular resonances
involving color singlet $q\bar{q}$ pairs.  Our results also agree with lattice
and other models predicting that color exotics (hybrid mesons, octet-octet,
sextet-sextet and triplet-triplet tetraquarks) will have masses near and above 2 GeV.
Finally, we have utilized the Frank Condon principle and the distinctive quark
momentum distribution in a tetraquark to propose a new off-plane correlator
measurement for identifying exotic hadrons.

{\bf Acknowledgments}

This work was supported in part by grants FPA 2004-02602, 2005-02327,
PR27/05-13955-BSCH (Spain) and U. S. DOE
DE-FG02-97ER41048 and DE-FG02-03ER41260.


\end{document}